\documentclass[namedreferences]{solarphysics}
\usepackage[hyperref,optionalrh]{spr-sola-addons} 
\usepackage{graphicx}        
\usepackage{color}           
\usepackage{breakurl}        

\chardef\us=`\_

\begin{document}

\begin{article}

\begin{opening}

\title{Automatic Detection of Occulted Hard X-ray Flares Using Deep-Learning Methods}

\author[addressref={aff1,aff2},corref,email={shinnosuke.ishikawa@rikkyo.ac.jp}]{\inits{S.}\fnm{Shin-nosuke}~\lnm{Ishikawa}\orcid{0000-0002-5125-9499}}
\author[addressref=aff3]{\inits{H.}\fnm{Hideaki}~\lnm{Matsumura}}
\author[addressref={aff2,aff3}]{\inits{Y.}\fnm{Yasunobu}~\lnm{Uchiyama}}
\author[addressref=aff4]{\inits{L.}\fnm{Lindsay}~\lnm{Glesener}\orcid{0000-0001-7092-2703}}
\address[id=aff1]{Strategic Digital Business Unit, Mamezou Co., Ltd., 2-1-1 Nishi-Shinjuku, Shinjuku, Tokyo 163-0434, Japan}
\address[id=aff2]{Graduate School of Artificial Intelligence and Science, Rikkyo University, 3-34-1 Nishi-Ikebukuro, Toshima, Tokyo 171-8501, Japan}
\address[id=aff3]{Galaxies Inc., 1-1-11 Minami-Ikebukuro, Toshima, Tokyo 171-0022, Japan}
\address[id=aff4]{School of Physics and Astronomy, University of Minnesota, Minneapolis, MN 55455, USA}

\runningauthor{S. Ishikawa et al.}
\runningtitle{Automatic Detection of Occulted Flares Using Deep Learning}

\begin{abstract}
We present a concept for a machine-learning classification of hard X-ray (HXR) emissions from solar flares observed by the \textit{Reuven Ramaty High Energy Solar Spectroscopic Imager} (RHESSI), identifying flares that are either occulted by the solar limb or located on the solar disk. 
Although HXR observations of occulted flares are important for particle-acceleration studies, HXR data analyses for past observations were time consuming and required specialized expertise. 
Machine-learning techniques are promising for this situation, and we constructed a sample model to demonstrate the concept using a deep-learning technique. 
Input data to the model are HXR spectrograms that are easily produced from RHESSI data.  The model can detect occulted flares without the need for image reconstruction nor for visual inspection by experts. 
A technique of convolutional neural networks was used in this model by regarding the input data as images. 
Our model achieved a classification accuracy better than 90\,\%, and the ability for the application of the method to either event screening or for an event alert for occulted flares was successfully demonstrated. 
\end{abstract}

%

\end{opening}

%

\section{Introduction}

Hard X-ray (HXR) emissions from solar flares provide important information on how particles are accelerated to high energies.
In particular, non-thermal HXR emissions from the solar corona are important because they are emitted near energy-release sites just after acceleration.
In many flares, HXR sources at footpoints of magnetic-loop structures are prominent compared to coronal sources \citep{krucker2008review}. 
Therefore, it is challenging to observe weak HXR sources in the corona, with stronger HXR sources that are relatively near.  

This problem requires a high imaging dynamic range, but 
the dynamic range is limited with past instruments for solar HXR observations. 
Unlike visible-light optics, it is technically difficult to focus HXRs using optics (mirrors and lenses), so indirect imaging methods are typically used.  Applications of directly focusing optics for solar HXR observations have been made but are so far limited to short-term observations \citep{krucker2014, christe2016, grefenstette2016}. 
However, to study particle acceleration, which is usually detected in medium to large flares, long-term observation is necessary. 
The most thorough available data set to date for solar HXRs comes from the \textit{Reuven Ramaty High Energy Solar Spectroscopic Imager} \citep[RHESSI:][]{lin2002} spacecraft, which operated from 2002 to 2018.
The imaging technique used by RHESSI is a rotating modulation collimator, and it is not optimized for high dynamic range observations \citep{hurford2002}.  

One of the ways to investigate coronal HXR sources with RHESSI data is to select flares whose footpoints are behind the solar limb.
In those flares, footpoint HXR emissions are absorbed by the lower layers of the Sun and are not visible to the observing instrument.  
In such a case, there is the possibility to more easily observe coronal HXR sources, which are at, or near, the places where particle acceleration is thought to occur. 
Those flares are often referred to as partially occulted flares, and are thought to be important events for investigating features of accelerated particles.  
In this article, we define a partially occulted flare, or just an ``occulted flare'', as a flare that occurred near the limb without a detection of a HXR source at its footpoint(s). 

\begin{figure} 
 \centerline{\includegraphics[width=1\textwidth,clip=]{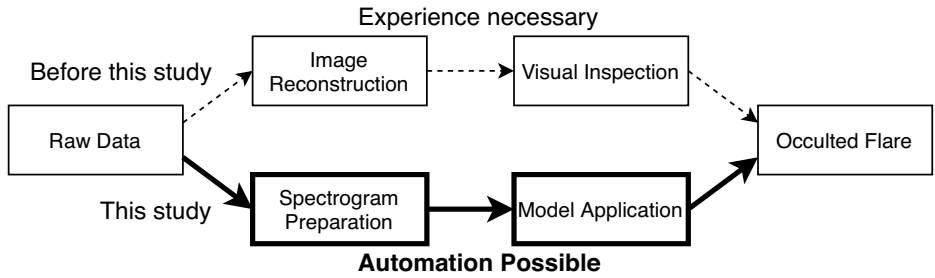}}
 \caption{Diagram comparing two methods of searching for occulted flares.  The upper path shows a method using manual image reconstruction and visual inspection by an experienced researcher.  
 The lower path shows the method suggested in this study using a machine-learning model.  }\label{fig:diagram}
 \end{figure}

Occulted flares observed by RHESSI have been published as catalogs.
\citet{krucker2008survey} studied 55 events, and \citet{effenberger2017} added to that catalog 61 more events that occurred later. 
Both studies identified the occulted flares by checking HXR images of candidate flares by eye for every event. 
To perform such a search, two major steps need to be performed by an experienced HXR imaging analyst (upper path in Figure~\ref{fig:diagram}): image reconstruction and visual inspection. 
Obtaining HXR images with good quality from the RHESSI data is not a trivial effort with modulation collimators.  
A complicated image-reconstruction method is necessary, and knowledge and experience are required to obtain reliable images.  
It is also difficult to determine which features of a reconstructed image are real and which are artificially generated due to noise in the reconstruction algorithm. 
As a result, statistical studies of occulted flares were time consuming and not available to non-experts.  
We note that the RHESSI mission archive (\url{https://hesperia.gsfc.nasa.gov/rhessi3/mission-archive/index.html}) has partially alleviated one of these problems, by providing automatically generated images in several energy ranges for \textbf{all} RHESSI flares.  However, identifying occulted flares using traditional methods would still require visual inspection of all of these events by an expert, so examining the entire RHESSI database in that way would be impossible.

A better situation would be if a broad range of researchers can determine whether flares are partially occulted or on the solar disk without spending too much time, and application of artificial intelligence (AI) technologies, especially machine-learning techniques, is a candidate to overcome that difficulty. 
By treating data without image reconstruction as explanatory variables, we can formulate this problem as a two-class classification problem.  
If we construct and train a machine-learning model with a reasonable accuracy, we can reduce the time and effort to find occulted events and focus more on scientific analysis of target events.

In the field of solar physics, several recent studies use machine-learning techniques for space weather prediction or image classification.
Specifically, deep learning (DL) using models of deep neural networks (DNNs) are used for complicated data sets including image data. 
\citet{nishizuka2018} and \citet{panos2020} each use DNNs for their solar flare prediction models.
A convolutional neural network (CNN) is a kind of DNN especially applied for image recognition, and it is 
used for classifying solar images in categories of filaments, flare ribbons, prominences, sunspots, and no feature regions \citep{armstrong2019}.
In addition, CNN is also used for instrumental calibration of solar telescopes \citep{neuberg2019}.
The generative adversarial network (GAN) technique is an application of DNN used for generating images from other data.  
GAN is used for generation of solar images from magnetograms \citep{park2019}, and solar image deconvolution \citep{xu2020}.

So far, in most of the classification studies using machine learning in solar physics, input data were images understandable to humans, such as UV images and magnetograms. 
While it is true that those automations are natural applications of machine-learning techniques, 
another advantage of machine-learning algorithms is the ability to judge data that humans cannot recognize with our eyes only. 
The input data are not necessarily in an understandable form.  
One example of non-understandable image classification is a malware detection algorithm \citep{nataraj2011}.
In this study, they made ``images'' from binary data files, and they classified whether each image contained malware or not based on the observed patterns.
Of course, humans cannot understand an image produced from binary data inside a data file, but we can use a CNN model since the data are represented as a two-dimensional image. 

In this study, we use RHESSI spectrograms as input data and present a concept of a CNN model to judge whether test flares are occulted or not. 
The RHESSI spectrograms can be systematically produced without specialized expertise, and we can search for occulted flares automatically without an expert and without the need for individual inspection (lower path in Figure~\ref{fig:diagram}). 
A major goal of this work is to liberate the extensive database of indirect solar HXR observations from the realm of specialists and make its scientific content more easily available.  This will induce more active researchers to study particle acceleration in the Sun. This effort supports and enhances the use of the RHESSI legacy mission archive.

\section{Data Set and Modeling}
\subsection{Data Format and Preparation}
To train and test a machine-learning model to detect occulted flares for supervised learning, we prepared HXR spectrograms observed by RHESSI during flares as input data.
A RHESSI spectrogram is a two-dimensional histogram with axes of time and energy, and the value in each bin indicating the number of counts observed by RHESSI at that time and energy. 
Since RHESSI's imaging technique works by modulating the photon flux, the intensity changes over time contain both temporal and positional information for the HXR source.
Imaging analyses are typically performed using data binned in intervals that are an integral multiple of the modulation period, which corresponds to the spacecraft rotation period of approximately four seconds. 
Rather than separating spectral and positional data, we use the spectrograms themselves as input data; we can then apply models that are well developed for image recognition with little modification. 

We defined an input data format as a 100~pixel $\times$ 100~pixel histogram covering one rotation period for the time axis ($\approx$0.04\,second bin size) and 2\,--\,52\,keV for the energy axis (0.5\,keV bin size). 
We set the time interval such that it covers one full rotation of the RHESSI grids, and the time resolution such that each bin contributes significantly to the modulation.  (See discussion of imaging timescales by \citet{lin2002}.) 
For the energy, we set the range to cover both thermal and non-thermal components, and the resolution to be slightly less than the instrumental spectral resolution.
We also attempted several other resolutions for temporal and energy bins, and we found no significant performance improvement with resolutions finer than 100~pixels $\times$ 100~pixels.

There are 116 occulted-flare events published by \citet{krucker2008survey} and \citet{effenberger2017} that occurred between 2002 and 2015, and we used all of these published events for the data set of occulted flares. 
Since 116 data are not enough to train a CNN model in most cases, we made ten spectrograms for each event using ten consecutive time intervals. 
In the context of machine learning, this is related to the concept of data augmentation to obtain enough training data from a limited sized data set. 
The time intervals are selected from the time range between five rotation periods before the flare peak time to five periods after it, and there is no time overlap between the intervals. 
The peak time values are from the RHESSI flare list (\url{https://hesperia.gsfc.nasa.gov/rhessi3/data-access/rhessi-data/flare-list/index.html}).
We therefore obtained 1160 spectrograms of flares that are known to be occulted.

For the data set of flares not occulted by the solar limb, we randomly selected 1000 events from the RHESSI flare list that had a radial distance from the solar disk center of $<300''$.  This ensures that this set of flares occurred on the disk.  
In addition, we selected only flares with a peak count rate of $>100$~counts\,s$^{-1}$ to ensure sufficient statistics for each spectrogram. 
We did not set a criterion on the time of the flare observation.  
One spectrogram for each event with a time interval starting from the peak time was produced, and 1000 spectrograms in total were used for the on-disk data set. 

\begin{figure}
 \centerline{\includegraphics[width=1\textwidth,clip=]{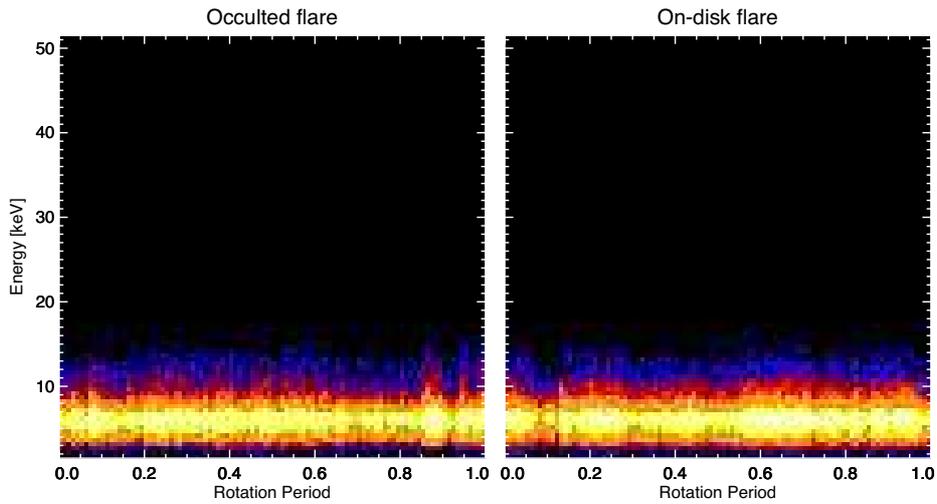}}
 \caption{Sample spectrograms for the input data shown with a log scale, including an example of (left) an occulted flare and (right) an on-disk flare. }\label{fig:spectrogram}
 \end{figure}

We added labels of either of two classes - ``occulted'' or ``on-disk'' - to all of the spectrograms. 
Examples of the spectrograms for the occulted and on-disk flares are shown in Figure~\ref{fig:spectrogram}.  
It is difficult to tell by visual inspection which features in the spectrograms correspond to the labels.  
The problem to be solved in this article is formulated to classify this kind of difficult-to-understand images, and we expect that the machine-learning model can distinguish them even though the human eye cannot.  

We split each of the occulted and on-disk data set into 80 and 20\,\% at random for training and testing the machine-learning model.  
24 occulted flares are picked and 240 spectrograms of those flares were used for the test data set.  
The other 920 spectrograms were used for the model training.  
We note that we did not use spectrograms from any of the same flares for both the training and test data sets. 
Spectrograms from the same flare are thought to have similar features, so there is a risk to overestimate the performance if we include them for both the training and test data sets.
800 and 200 randomly sampled on-disk flare spectrograms were used for the training and test sets.

\subsection{Model}
We constructed the DL model with CNNs based on the residual net \citep[ResNet:][]{he2015} model originally developed to classify a large set of images (ImageNet).
A feature of the ResNet is a concept to train only residuals of each layer of neural networks against the input data for better training of neural networks deeper than before. 
Accuracy of ImageNet data classification with a machine-learning model outperformed that of a human for the first time by \citet{he2015b} with a new training parameter initializing method (known as He initialization), and the same group 
improved the accuracy further with the ResNet models in the same year. 

\begin{figure}
 \centerline{\includegraphics[width=1\textwidth,clip=]{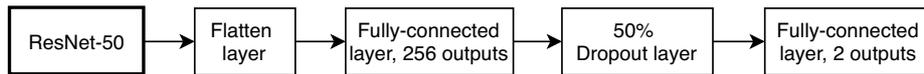}}
 \caption{The DL model presented in this article based on the ResNet-50 model \citep{he2015}.}\label{fig:modeldiagram}
 \end{figure}

We added a few layers to a 50 layer ResNet model (ResNet-50) to make a two-class classifier. 
A flatten layer was added to convert the ResNet-50 output to a one-dimensional array.  
Two fully connected layers with 256 and 2 outputs were added to reduce the output size gradually, and a dropout layer \citep{srivastava2014} was added between the fully connected layers to avoid overfitting. 
The model configuration is summarized in Figure~\ref{fig:modeldiagram}.  This model has 31,923,970 trainable parameters. 
We trained the model with cross entropy loss function and Adam optimizer \citep{kingma2014} with a learning rate of $3 \times 10^5$.
Mini-batch training was performed with a batch size of 64. 

We implemented the model using the machine-learning framework TensorFlow version 1.14.0 \citep{abadi2015} 
in combination with the neural network wrapper library Keras version 2.2.4 \citep{chollet2015}.
The model construction, training, and test are performed using Anaconda3 5.3.0 with Python version 3.6.8.
The ResNet-50 model with the trained parameter set by ImageNet is available with the Keras library, and we utilized it as the starting point for building our model.

\section{Result}
We trained the CNN model for the flare event classification using the training data described in the previous section.  
Initial parameters for the ResNet-50 part of the model were the parameters obtained by training ImageNet.  
The parameters are for color images with red, green, and blue channels, and we put identical two-dimensional spectrogram data for all three color channels of the model input.  
This concept corresponds to transfer learning or fine tuning of an existing trained machine-learning model. 
Although it is usually used for new data sets with similarities with the original data used for the model training, 
we found the model training with the HXR spectrograms got better results initializing with the ImageNet parameters than when not. 
The ImageNet parameters are only initial values; we did not fix any parameters from them.
Even so, the performance was better than using random initialization.
The trained model is available as electronic supplementary material and in the Data Repository for University of Minnesota: \url{{https://doi.org/10.13020/wtbm-2258}}

We then tested the model with the test data set described in the previous section.  The model classified correctly for $\gtrsim$90\,\% of the cases.  
We made several trials with different splits for training and test data sets with the same optimizer, loss function, and hyper-parameters, and the results ranged within a few percent.  
We show detailed results for one of these trials here.  
In this trial, 226 out of the 240 occulted flare test data were successfully categorized as occulted.  
For the on-disk events, 179 out of the 200 data were successfully categorized. 
Focusing on the occulted-event detection, the true positive (TP) and true negative (TN) cases are the 226 and 179 events. 
The accuracy is defined by 
\begin{eqnarray}
 \mbox{Accuracy}=\frac{\mbox{TP}+\mbox{TN}}{\mbox{[Total Cases]}},
\end{eqnarray}
and it is calculated to be 92\,\%.
Although the accuracy varied depending on training and test data set selections in the several trials, 
it was $>$90\,\% for most cases and reached a maximum of 94\,\%. 

The precision and recall, which are defined as 
\begin{eqnarray}
 \mbox{Precision}=\frac{\mbox{TP}}{\mbox{TP}+\mbox{FP}}, \\
 \mbox{Recall}=\frac{\mbox{TP}}{\mbox{TP}+\mbox{FN}},
\end{eqnarray}
where FP and FN are false positive and false negative cases, were 91\,\% and 94\,\% for the occulted events in the trial shown in the previous paragraph. 
The precision corresponds to the ratio of the actually occulted events out of the events classified as occulted, 
 and the recall corresponds to the coverage of the occulted event detections out of the total occulted events. 
The result of those measures are similarly high, and it means we do not have many FN cases and we do not miss many occulted flares.  
The $F_1$ score, which is the harmonic mean of the precision and recall described as 
\begin{eqnarray}
 [F_1\mbox{-score]}=\frac{2 \mbox{[Precision][Recall]}}{ \mbox{[Precision]}+\mbox{[Recall]}}=\frac{2\mbox{TP}}{2\mbox{TP}+\mbox{FP}+\mbox{FN}}, 
\end{eqnarray}
was calculated to be 92\,\% for the occulted flares.  
Therefore, it is confirmed that the model has an ability to detect occulted flares from systematically generated HXR spectrograms. 
The statistical measures are summarized in Table~\ref{table:measures} for the occulted and on-disk events.
\begin{table}
\caption{Summary of the statistical measures for one model test.  The total accuracy was 92\,\%. }
\label{table:measures}
\begin{tabular}{cccc}  
  \hline
Label & Precision [\%] & Recall [\%] & $F_1$ score [\%] \\
  \hline
Occulted & 91 & 94 & 92\\
On-disk & 93 & 89 & 91\\
  \hline
\end{tabular}
\end{table}

Since the objective of this work is to demonstrate the concept of automatic detection of occulted flares without specialized expertise, 
we purposefully did not fine-tune the model. 
The classification accuracy could be much higher if hyper-parameter tuning and/or model architecture modification were made.  
We found that the ResNet-50 model achieved better performance than other models based on built-in models in the Keras library. 
We did adjust some hyper-parameters such as learning rate and batch size to achieve better results with this model.

Not much computer resources were necessary to train this model.  
We performed the model training not with a GPU server but with a CPU in a laptop computer (MacBook 2017 model with 1.4\,GHz dual core Intel Core i7 and 16\,GB memory).
It took less $\approx$300\,seconds per epoch, and we trained the model for 10 epochs since initial tests did not show any improvement after 10 epochs. 
Applying the classification using the already-trained model is even easier;  
one can apply this model to a set of spectrograms in much less than an hour with a laptop CPU. 
With our MacBook, it took less than a minute to classify the 648 events described in the following section. 

\section{Discussion}
The accuracy of $\approx 90$\,\% or better for the occulted versus on-disk flare classification shows 
that this technique could be useful for event screening in the archive data, or for future observations. 
The accuracy from this initial demonstration is not high enough to identify \textit{all} of the occulted flares from the whole RHESSI flare list, but it would identify the vast majority ($\approx 90$\,\%) of them, providing plenty of events for statistical studies. 
We can utilize the model in combination with other information such as radial distance from the Sun center, flare size, and energy range detected. 
If the radial distance is small, the probability of misdetection is high even if it is classified as occulted by the model.  But these flares are easy to exclude from the analysis and would not affect studies for occulted flares. 
If the energy range detected for a flare only included thermal emission (such as up to $\approx$12~keV or less) and not higher-energy, nonthermal emission, or the flare was faint, 
the probability of a misdetection is higher.  But these flares are not often the ones desired to study for particle acceleration.  
Therefore, one usage for this model is to select the set of events to analyze in detail.    
We would obtain a reasonable number of events by using additional criteria along with the classification result, such as basic information from the flare list, and we can take more time for scientific analyses for more suspect events. 
We note that the model presented in this article was trained with the on-disk events with $<300''$ from the disk center, and there is a possibility that the performance to classify events close to the limb improves if we train the model using the on-disk events closer to the limb.  
In addition to that, it is also possible to use the model to perform an onboard event alert for occulted flares for future observations (e.g. \textit{Solar Orbiter}, in which other instruments could respond to a flare trigger by the hard X-ray instrument).  

We have tested the model for events from the RHESSI flare catalog with criteria of a peak count rate to be $>$100~counts\,s$^{-1}$, 
a radial distance to be $>$950$''$ and detection energy range to be up to 25\,keV or more. 
648 events meet these criteria, and 536 events are  classified as occulted events by our model. 
The 648 events include 78 occulted events from the lists published by \citet{krucker2008survey} and \citet{effenberger2017}, and 76 of those events were successfully classified as occulted events. 
In addition to these are the other 460 events classified as occulted but not on the previously published lists. 
These include the rather famous X-class flare that occurred on September 10 2017, and we can confirm that the HXR footpoint sources were actually occulted over the solar limb from previously published results \citep{gary2018, ovchinnikova2019}. 
As a result, we successfully demonstrated the ability to find an occulted event not in our training or test data sets.

In general, for standard DL models, it is difficult to interpret which features of the input data are important for the result.   
The is true for the model presented in this article, and we cannot tell which feature in the HXR spectrograms is important for identifying the occulted flares.  We can guess at possibilities for the differences in HXR spectrograms of occulted and on-disk flares. In general, intensities of coronal HXR sources are weaker than those of footpoint HXR sources, and source areas are generally larger for coronal sources. 
In HXR observations with both coronal and footpoint source detections, coronal sources tend to have steeper spectral indices \citep{masuda1995, ishikawa2011looptop, krucker2014}. 
This might make a difference in spectra for the occulted flares, which show only a coronal source in the HXR range, in contrast to the on-disk flares with strong footpoint sources. Future work will examine spatial and spectral characteristics of the false positives in order to determine whether such trends are important for the DL model identification..  
As described in the previous section, we did not consume too much time in optimizing the model to achieve the highest possible accuracy.  
This means that there remains potential for future work to construct an optimized model that can detect the occulted flares with even better accuracy. 
Even in that case, we note that this model should be used primarily for event screening, not for directly deriving scientific results.

In this article, we focused only on a two-class classification of the HXR sources, i.e. whether they are occulted or on-disk. 
For future work, classification in more than two classes would be possible based on the successful result of this model.
Also, a model to classify, detect, or predict features of solar phenomena would be possible using a combination of observations in HXR and other wavelengths. 

\section{Summary}
We constructed a DL model to classify HXR observations of solar flares as occulted or on-disk, and have demonstrated a successful concept to detect occulted HXR flares automatically without too much time or presence of an imaging analysis expert. 
An accuracy of $\approx 90$\,\% was achieved for the test data set, and successfully demonstrated the ability to find occulted flares by detecting another flare not on the existing list of occulted flares. 
We can directly apply this model for event screening, and there exists a wide range of application for the DL technique in this scientific area.

%

%
\begin{acks}
 We would like to thank Tomoe Hoshi for introducing an example of non-understandable image analysis.
\end{acks}

{\footnotesize\paragraph*{Disclosure of Potential Conflicts of Interest}
The authors declare that they have no conflicts of interest. 
}

%
%
 \bibliographystyle{spr-mp-sola}
 \bibliography{ishikawa2021.bib}  
%

%
%
%

\end{article} 
\end{document}